\documentclass[preprint]{elsarticle}

\usepackage{amssymb, latexsym, amsthm, amsmath, lineno, epsfig, mathtools, hyperref, todonotes, booktabs, cite, graphicx}
\usepackage[singlelinecheck=false]{caption}
\modulolinenumbers[1]

\journal{Theoretical Population Biology}

\newcommand{\bs}[1]{\ensuremath{\boldsymbol{#1}}}

\newcommand\E{\operatorname{E}}

\newcommand\given{{\,|\,}}

\newcommand\eg{{\it e.g.,}}

\newcommand\ie{{\it i.e.,}}

\newcommand\x[1]{\ensuremath{x_{#1}}}
\newcommand\y{\ensuremath{y}}
\newcommand\s{\ensuremath{s}}
\newcommand\fv[1]{\ensuremath{\mathbf{f}_{#1}}}
\newcommand\bv[1]{\ensuremath{\mathbf{b}_{#1}}}

\begin{document}

\begin{frontmatter}

\title{Inference of population genetic parameters with a biallelic mutation drift model using the coalescent, diffusion with orthogonal polynomials, and the Moran model}

\author[address1,address2]{Claus Vogl\corref{correspondingauthor}}
\cortext[correspondingauthor]{Corresponding author}
\ead{claus.vogl@vetmeduni.ac.at}
\author[address1,address2]{Sandra Peer}
\ead{sandra.peer@gmail.com}

\address[address1]{Department of Biomedical Sciences, Vetmeduni Vienna, Veterin\"arplatz 1, A-1210 Wien, Austria}
\address[address2]{Vienna Graduate School of Population Genetics, A-1210 Wien, Austria}

\begin{abstract}
In population genetics, extant samples are usually used for inference of past population genetic forces. With the Kingman coalescent and the backward diffusion equation, inference of the marginal likelihood proceeds from an extant sample backward in time. Conditional on an extant sample, the Moran model can also be used backward in time with identical results, up to a scaling of time. In particular, all three approaches---the coalescent, the backward diffusion, and the Moran model---lead to the identical marginal likelihood of the sample. If probabilities of ancestral states are also inferred, either of discrete ancestral allele particle configurations, as in the coalescent, or of ancestral population allele proportions, as in the backward diffusion, the backward algorithm needs to be combined with the corresponding forward algorithm to the forward-backward algorithm. Generally orthogonal polynomials, solving the diffusion equation, are numerically simpler than the other approaches: they implicitly sum over many intermediate ancestral particle configurations; furthermore, while the Moran model requires iterative matrix multiplication with a transition matrix of a dimension of the population size squared, expansion of the polynomials is only necessary up to the sample size. For discrete samples, forward-in-time moving pure birth processes similar to the Polya- or Hoppe-urn models complement the backward-looking coalescent. Because, the sample size is a random variable forward in time, pure-birth processes are unsuited to model population demography given extant samples. With orthogonal polynomials, however, not only ancestral allele proportions but also probabilities of ancestral particle configurations can be calculated easily. Assuming only mutation and drift, the use of orthogonal polynomials is numerically advantageous over alternative strategies.
\end{abstract}


\end{frontmatter}


\section{Introduction}

In population genetics, present-time (extant) data are usually used for inference of past population genetic processes and forces. The coalescent \citep{King82} is a stochastic process that describes the genealogy of a sample from a single locus backward in time, until the last common ancestor of the sample is reached. It allows for convenient simulation of a genealogical tree, conditional on the current sample size and the (effective) population size or sizes in the past. Subsequently, a mutation process may distribute allele states on this tree. The coalescent has become a pillar of population genetics, described in easy-to-read textbooks \citep[\eg][]{Hein05,Wake09} and implemented in easy-to-use software packages, \eg\ {\it ms} \citep{Huds90}. 

With the coalescent process, the extant sample is given but, at any time in the past, generally many intermediate configurations are possible until the last common ancestor is reached. Mutations increase this complexity further. Coalescent simulations have been used for inferring parameters of relatively complex population genetic models using high-throughput population genetic data. Because of the combinatorial complexity of the coalescent with mutation, sufficient statistics are often not available, such that summary statistics and Approximate Bayesian Computation (ABC) \citep[\eg][]{Beau02,Esto10,Frai17} are used for inference of population genetic parameters. Asides from wasting information, because the summary statistics are generally not sufficient, and thus allowing only for approximate inference, ABC is also computationally demanding, such that often only subsets of the parameter space can be investigated. 

Earlier, the infinite sites model was often used assuming mutation-drift equilibrium and complete linkage disequilibrium, \ie\  no recombination \citep[\eg][]{Watt75}. In population genetic model species of the genus {\it Drosophila,} deviations from linkage equilibrium are barely noticeable in genomic regions of moderate to high recombination rates. The numbers of mutations in a short intronic region are only very slightly more dispersed than a binomial distribution, where independence is assumed \citep{Clem12a}, indicating that the effects of linkage are at most weak. Furthermore, this overdispersion may be caused by forces other than linkage, such as fluctuations in the effective population size or mutation rate over the genome. Thus linkage can presumably be neglected. On the other hand, deviations from mutation-drift equilibrium are readily apparent and should not be ignored \citep[\eg][]{Clem12a}. Furthermore, only extremely rarely more than two alleles segregate in a moderately-sized sample at a nucleotide site. Thus a bi-allelic mutation model seems sufficient. Assuming linkage equilibrium allows compact representation of data in a site frequency spectrum (SFS), for a single population, or a joint site frequency spectrum (jSFS), for multiple populations. While linkage equilibrium and bi-allelic loci may thus reasonably be assumed, deviations from mutation-drift equilibrium should be expected; in particular, effective population sizes may change noticeably. 

The bi-allelic mutation model can be re-parametrized to a parent-inde\-pen\-dent mutation model. This simplifies theory as, both with a mutation and a coalescent event, the ancestral state before the event is irrelevant. Abandoning the separation of the coalescent from the mutation process, simple particle models can then be derived. Forward in time, these models resemble Polya's urn: with a parent-independent mutation model, a pure-birth process results \citep{Donn96a,Step00}. Combining forward- and backward-running particle models, {\bf ancestral particle configurations} conditional on the sample's current state can be inferred. On the other hand, the {\bf ancestral allele proportion} conditional on the sample's current state can be inferred using orthogonal polynomials that solve the forward or backward diffusion model; in the case of the bi-allelic mutation drift model, the modified Jacobi polynomials are appropriate \citep{Song12}. With both particle and continuous approaches, the forward-backward algorithm (a dynamic programming approach) can be used for relatively simple and efficient calculation \citep[\eg][]{Berg17}. 

Particle models as well as orthogonal polynomials allow for inference of population genetic parameters. Marginalization is straightforward: with particle models, ancestral particle configurations can be ``summed out''; with ortho\-gonal polynomials, the ancestral allele proportion can be ``integrated out'' \citep{Berg17}. Crucially, the ortho\-gonal polynomial approach does not suffer from the combinatorial complexity of the particle approach. It therefore allows for exact inference with relatively complex population genetic models, while particle approaches allow for exact inference only in the simplest cases. With ortho\-gonal polynomials, both forward- and backward-in-time algorithms can be used outside of equilibrium, when \eg\ a changing population size is modeled \citep{Berg17}. With the particle approach, however, changing population size can only be accommodated easily backward in time, since then conditioning on the extant sample size is possible, whereas the sample size becomes a random variable forward in time. Even so, forward particle probabilities are necessary to infer the probability of an ancestral configuration some time in the past, conditional on the extant sample configuration. 

The particle and the orthogonal polynomial approaches are closely connected \citep{Ethe09}. Indeed, the two approaches are dual, such that the expectation with respect to the population allele frequencies with orthogonal polynomials is identical to the expectation with respect to particle configurations in the particle approach \citep{Ethe09,Grif10}.---Unfortunately, this theory has not been formulated for ease of inference or implementation on a computer. 

In this article, we start from the haploid Moran model with population size $N$ and the equations for coalescing particles assuming bi-allelic mutation, forward and backward in time. As long as the sample size $M$ is at most equal to the population size $N$ and time is scaled appropriately, the coalescent is consistent, such that the backward-in-time coalescent-mutation pure-death process proceeds identically, irrespective of the population size $N$ \citep[\eg][]{Ethe11}. This extends also to the diffusion limit $N\to\infty$. While usually lineages are traced back to the last common ancestor, which corresponds to a sample of size $m=1$, we trace the sample size back to $m=0$ to show the equivalence to orthogonal polynomials. Another reason for extending times until samples of size zero are reached is that the approach can also be used for inference in phylogenetic contexts \citep{DeMa13,Schrempf2016, Berg17}, where species split-times can be much longer than coalescent and mutation rates. Furthermore, the connections to Polya's urn are even more evident then.

While results from particle models and orthogonal polynomials are identical, as long as time is scaled appropriately, inference with orthogonal polynomials is numerically simpler, as orthogonal polynomials are eigenfunctions of the forward and backward diffusion equations, and thus preferable. This allows efficient calculation of the marginal likelihood and thus inference of population parameters, which is usually all population geneticists are interested in. (We note that, in mutation-drift equilibrium, efficient inference methods for both the scaled mutation rate and the mutation bias are available: an expectation-maximization algorithm for maximum likelihood inference and a Markov-chain Monte Carlo method for inference of the posterior distribution \citep{Vogl14b}.) Orthogonal polynomials also implicitly sum over particle configurations. Furthermore, simple linear transformations from regular polynomials to orthogonal polynomials and vice-versa are available. Thus, it is easily possible to obtain the moments of the distribution \citep[compare][]{Evan07}, given orthogonal polynomials. Orthogonal polynomials may be interpreted as linear combinations of particle probabilities assuming a Moran model with a finite population size $N\geq M$. In a second step, probabilities of the ancestral particle configurations, which are usually inferred with particle models, may also be obtained conveniently as linear functions of orthogonal polynomials. This allows calculation of probabilities, even out of equilibrium, \eg\ if effective population sizes change. Generally, we provide powerful and efficient algorithms to compute particle probabilities using orthogonal polynomials. 


\section{Population allele proportions in the Moran and diffusion model}

In this section, we review the bi-allelic version of the Moran and diffusion model equations and the equations for joint and conditional probabilities, given a sample from the present. 

\subsection{The Moran and diffusion model equations}

Consider a bi-allelic reversible decoupled mutation-drift Moran model with mutation rates per event of $\mu_1$ towards and $\mu_0$ from the focal allele with haploid population size $N$. We re-parametrize by setting the mutation bias towards the focal allele to $\alpha=\mu_1/(\mu_0+\mu_1)$ with $0<\alpha<1$ and introducing the total mutation rate $\mu=\mu_0+\mu_1$. Furthermore, we introduce $\beta=1-\alpha$, for simplicity of notation. Let $i$ ($0\leq i\leq N$) be the frequency of allele one. Then, the tri-diagonal transition rate matrix $\mathbf{T}$ of allele proportions $\x{t}$ to $\x{t+1}$ depends on $N$, $\mu$ and $\alpha$, with transition probabilities
\begin{equation}\label{eq:transition_decoupled_Moran}
\begin{cases}
\Pr(\x{t+1}=\frac{i-1}N\given \x{t}=\frac{i}{N})&=\x{t}(1-\x{t})+\beta\mu \x{t}\\
    \Pr(\x{t+1}=\frac{i}{N}\given \x{t}=\frac{i}{N})&=1-2 \x{t}(1-\x{t})-\beta\mu \x{t} - \alpha\mu \x{t}\\
\Pr(\x{t+1}=\frac{i+1}N\given \x{t}=\frac{i}{N})&=\x{t}(1-\x{t})+\alpha\mu \x{t}\,.
\end{cases}
\end{equation}
If exponential waiting times between Moran reproduction events are also incorporated, the Moran model becomes continuous in time. In particular, setting the rate parameter for reproduction events of the Moran model to $N^2$ and the overall scaled mutation rate to $\theta=N(\mu_0+\mu_1)$ makes them equivalent to those of the diffusion model below. Setting $\delta x=1/N$, the forward operator of the continuous-time $N$-particle Markov process \citep{Ethe11} is
\begin{equation}
    \frac{d}{dt}\phi(x\given t) =
    {\cal L}_N\phi(x\given t)
\end{equation}
with the infinitesimal generator
\begin{equation}\label{eq:forw_partly_cont_mutation}
\begin{split}
{\cal L}_N\phi(x\given t) &=
\alpha\theta \bigg(\frac{(1-x+\delta x)\phi(x-\delta x\given t) - (1-x)\phi(x\given t)}{\delta x}\bigg)\\
&\qquad+\beta\theta \bigg(\frac{(x+\delta x)\phi(x+\delta x\given t) - x\phi(x\given t)}{\delta x}\bigg)\\
&\qquad+\bigg(\frac{(x-\delta x)(1-x+\delta x)\phi(x-\delta x\given t)}{\delta x^2}\\
&\qquad+ \frac{(x+\delta x)(1-x-\delta x)\phi(x+\delta x\given t)}{\delta x^2}-\frac{2x(1-x)\phi(x\given t)}{\delta x^2}\bigg).\\
\end{split}
\end{equation}
where the first and second terms with $\theta$ correspond to mutation and the third term to genetic drift. 

We are concerned with samples of size $M$ from the population. With the discrete Moran model, a hypergeometric distribution is taken as likelihood, conditional on the population allele frequency $i=xN$ and population size $N$. This requires the sample size to be at most equal to the population size, \ie\  $M\leq N$. Then the likelihood of a frequency $\y$ of allele one in the sample is $\Pr(\y\given M,N,\x{t=0}N)$.

The backward generator corresponding to the forward generator (\ref{eq:forw_partly_cont_mutation}) is
\begin{equation}\label{eq:backw_partly_cont_mutation}
\begin{split}
&{\cal L}_N^{'}\Pr(\y\given M,x=\tfrac{i}{N}, t) = \\
&\qquad \frac{\alpha \theta(1-x) }{\delta x} \bigg(\Pr(\y\given M, x=\tfrac{i+1}{N}, t)-\Pr(y\given M, x=\tfrac{i}{N}, t)\bigg)\\
&\qquad+ \frac{\beta \theta x}{\delta x} \bigg(\Pr(\y\given M, x=\tfrac{i-1}{N}, t)-\Pr(\y\given M, x=\tfrac{i}{N}, t)\bigg)\\
&\qquad+ \frac{x(1-x)}{\delta x^2} \bigg(\Pr(\y\given M, x=\tfrac{i+1}{N}, t)+\Pr(\y\given M, x=\tfrac{i-1}{N}, t)\\
&\qquad\quad-2\Pr(\y\given M, x=\tfrac{i}{N}, t)\bigg)\,.
\end{split}
\end{equation}

The forward diffusion (Fokker-Planck) equation corresponding to the forward operator (\ref{eq:forw_partly_cont_mutation}) can be obtained by letting $N\to\infty$. Using the definitions of the first and second symmetric derivative, the operator of the forward diffusion (Fokker-Planck) equation is obtained from (\ref{eq:forw_partly_cont_mutation}) as
\begin{equation}\label{eq:forw_operator}
{\cal L} = -\frac{\partial}{\partial x}\theta(\alpha-x)\phi(x\given t) +\frac{\partial^2}{\partial x^2}x(1-x)\phi(x\given t),
\end{equation}
such that 
\begin{equation}\label{eq:forw_mutdrift}
    \frac{\partial}{\partial t}\phi(x\given t)={\cal L} \phi(x\given t)\,.
\end{equation}
Note that $\phi(x\given t)$ is a density. The corresponding operator of the backward diffusion equation obtained from  (\ref{eq:backw_partly_cont_mutation}) is
\begin{equation}\label{eq:backw_operator}
 {\cal L}^{'}=
    \theta(\alpha-x)\frac{\partial}{\partial x}\Pr(\y\given M, x, t)+x(1-x)\frac{\partial^2}{\partial x^2}\Pr(\y\given M, x, t)\,.
\end{equation}
With the continuous diffusion model, a binomial distribution with sample size $0\leq M<\infty$ is taken as likelihood conditional on the population allele proportion $x$ at time $t=0$, \ie\  $\Pr(y\given M, x,t=0)$. The corresponding backward diffusion equation is
\begin{equation}\label{eq:backw_mutdrift}
-\frac{\partial}{\partial t} \Pr(\y\given M, x,t)={\cal L}^{'}\Pr(\y\given M, x,t)\,.
\end{equation}
$\Pr(y\given M, x,t)$, a discrete probability distribution, is interpreted as the probability of obtaining a sample $(y,M)$ at the present, conditional on the allele proportion $x$ at time $t$ in the past \citep{Berg17}. The minus sign on the left side of the backward diffusion equation (\ref{eq:backw_mutdrift}) may be unusual \citep[compare][]{Ewen04}, but necessary such that the direction of time is compatible between the forward and backward diffusion models \citep{Zhao13a}. 

\subsection{Discrete Moran model: forward and backward algorithm}

With the Moran model, the forward-in-time starting distribution is a vector of probabilities $\bs{\rho}(x)$ at time $t=s$. 

Using the direction backward in time, we define the initial state
$\bv{0}$ as the hypergeometric likelihood $\bv{0}=\Pr(y\given M,x,N)$ at time $t=0$, 
\begin{equation}\label{hypergeometric}
    \Pr(y=i\given M=m,x,N)=\frac{\binom{x}{i}\binom{N-x}{m-i}}
    {\binom{N}{m}}\,,
\end{equation}
and recurse backward \citep{Rabi86,Vogl10,Berg17}
\begin{equation}
\begin{split}
\bv{t}' = \mathbf{T} \bv{t+1}' \quad (\s \le t <0)\,,
\end{split}
\end{equation}
where $\mathbf{T}$ is the discrete Moran transition matrix. Eventually, we obtain the marginal likelihood:
\begin{equation}\label{eq:marg_lh_alternative}
\begin{split}
\Pr(\y \given M,\alpha,\theta,\bs{\rho}) &= \bs{\rho} \mathbf{T}^{|\s|} \bv{0}'\,,
\end{split}
\end{equation}
where $\bv{0}$ is the hypergeometric likelihood $\Pr(y\given M,x,N)$ at time $t=0$ and $\mathbf{T}$ is the discrete Moran transition matrix. Forward in time, we start from the prior distribution and set $\fv{\s} = \bs{\rho}$, which corresponds to the vector of initial probabilities of states. We define recursively: 
\begin{equation}
\fv{t+1} = \fv{t}\mathbf{T} \quad (\s \le t < 0).
\end{equation}
Thus, $\fv{t}$ can be interpreted as the probability of the allele proportion at time $t$, conditional on the ancestral state $\bs{\rho}$, $\fv{t}=\Pr(\x{t}\given \bs{\rho})$. 

\paragraph{Joint and conditional probabilities} The probability of $\x{t} = i/N$ and $\y$ conditional on the starting distribution $\bs{\rho}$ is \begin{equation}\label{eq:joint_xy_discr}
\Pr(\x{t}=\tfrac{i}{N},\y \given \bs{\rho}) = (\fv{t})_i (\bv{t})_i\,.
\end{equation}
Furthermore, the probability of $\x{t} = i/N$ conditional on the data and the starting distribution is
\begin{equation}\label{eq:cond_x|y_discr}
\Pr(\x{t}=\tfrac{i}{N} \given \y,\bs{\rho}) = \frac{(\fv{t})_i (\bv{t})_i}{\fv{t}\bv{t}'}\,.
\end{equation}
This combination of forward- and backward-in-time calculations, \ie\  the forward-backward algorithm, allows calculation of the distribution of the population allele proportion conditional on the data and an initial condition $\bs{\rho}$ at any time $t$ \citep{Berg17}. 

\subsection{Diffusion model: forward and backward algorithm}

For the diffusion models, we follow \citet{Berg17} \citep[see also][]{Song12} and introduce the (modified) Jacobi polynomials (compare formula~22.3.2 in \citet{Abra70})
\begin{equation}\label{eq:Jacobi_modified}
  R_m^{(\alpha,\theta)}(x)=\sum_{l=0}^m(-1)^l\frac{\Gamma(m-1+l+\theta)\Gamma(m+\alpha\theta)}{\Gamma(m-1+\theta)\Gamma(l+\alpha\theta)l!(m-l)!}x^l\,.
\end{equation}
For $R_m^{(\alpha,\theta)}(x)$ the eigenvectors of the backward operator are
\begin{equation}
    -\lambda_n R_m^{(\alpha,\theta)}(x)={\cal L}^{'}R_m^{(\alpha,\theta)}(x)\,,
\end{equation}
with corresponding eigenvalues 
\begin{equation}
    \lambda_m=m(m+\theta-1)\,.
\end{equation}
The modified Jacobi polynomials fulfill the orthogonality relationship
\begin{equation}\label{eq:ortho_Jacobi}
    \int_0^1 R_n^{(\alpha,\theta)}(x) R_m^{(\alpha,\theta)}(x)\, w(x,\alpha,\theta)\,dx=\delta_{n,m} \Delta_n^{(\alpha,\theta)}\,,
\end{equation}
where $\delta_{n,m}$ is the Kronecker delta, 
$w(x,\alpha,\theta)=x^{\alpha\theta-1}(1-x)^{\beta\theta-1}$, and 
\begin{equation}
    \Delta_n^{(\alpha,\theta)}=\frac{\Gamma(n+\alpha\theta)\Gamma(n+\beta\theta)}{(2n+\theta-1)\Gamma(n+\theta-1)\Gamma(n+1)}\,.
\end{equation}

The binomial likelihood
\begin{equation}\label{eq:binomial}
    \Pr(y=i\given M=m,x,t=0)=\binom{m}{i}x^i(1-x)^{m-i}
\end{equation}
can be expressed uniquely by an expansion in the Jacobi polynomials up to order $M$ 
\begin{equation}\label{eq:binom2jacobi}
    \Pr(y\given M,x,t=0)=\binom{M}{y}x^y(1-x)^{M-y}=\sum_{m=0}^M d_m(M,y) R_m^{(\alpha,\theta)}(x)\,.
\end{equation}
Backward-in-time at $t$ ($s \le t \le 0$), we then have
\begin{equation}
    \Pr(y\given M,x,t)=\sum_{m=0}^M d_m(M,y) R_m^{(\alpha,\theta)}(x)e^{\lambda_m t}\,.
\end{equation}
The prior distribution $\rho(x)$ can also be expanded similarly with, possibly infinitely many, coefficients $\rho_n$. Then the continuous equivalent to the discrete marginal likelihood (\ref{eq:marg_lh_alternative}) is
\begin{equation} \label{eq:marg_like_general}
\begin{split}
\Pr(\y\given\rho)&=
    \int_0^1 \sum_{m=0}^M \rho_m d_m(M,y) w(x, \alpha, \theta) \big[R_m^{(\alpha,\theta)}(x)\big]^2 e^{\lambda_m s}\,dx\\
    &=\sum_{m=0}^M \rho_m d_m(M,y) \Delta_m^{(\alpha,\theta)} e^{\lambda_m s}\,. 
\end{split}
\end{equation}
Note that even though the expansion $\rho_n$ may be infinite, calculation of the marginal likelihood requires an expansion of only the order of the sample size $M$. 

Forward-in-time, set the initial distribution  $\phi(x\given t=\s)=\rho(x)$. At later time-points $t$, $\phi(x\given t,\rho)$ is calculated using the forward diffusion equation~(\ref{eq:forw_mutdrift}). The solution to the forward equation can then be represented as
\begin{equation}
\phi(x\given t,\rho)=w(x, \alpha, \theta) \sum_{m=0}^\infty \rho_m R_m^{(\alpha,\theta)}(x) e^{\lambda_m (s-t)}\,.
\end{equation}

\paragraph{Joint and conditional distributions}
The function corresponding to the joint distribution of the allelic proportion $x$ and the sample allele frequency $\y$ in the discrete case (\ref{eq:joint_xy_discr}) at time $t$ ($s \le t \le 0$) is
\begin{equation}\label{eq:joint_x_y}
\begin{split}
    j(x,\y \given t)&= \phi(x\given t,\rho)\Pr(y\given M,x,t)\\
    &=\sum_{n=0}^\infty \rho_n R_n^{(\alpha,\theta)}(x) e^{\lambda_n (s-t)}
    \sum_{m=0}^M d_m(M,y) R_m^{(\alpha,\theta)}(x)
    e^{\lambda_m t}\,.
\end{split}
\end{equation}
For the conditional distribution of the allele proportion $x$ given the sample allele frequency $y$, corresponding to eq.~(\ref{eq:cond_x|y_discr}) in the discrete case, $j(x,\y \given t)$ must be divided by the marginal likelihood (\ref{eq:marg_like_general})
\begin{equation}\label{eq:cond_x|y}
p(x\given \y,t,\rho)= \frac{j(x,\y\given t)}{\Pr(y\given \rho)}\,.
\end{equation}
Hence, both with the Moran and the diffusion models, the forward and backward algorithms can be used to infer the marginal likelihood of a sample conditional on a prior distribution some time $t=s$ in the past \citep{Berg17}. Since the coalescent is consistent \citep{Ethe11}, the results are identical, up to a scaling of time.

\section{Population allele proportions in particle models}

In this section, the transition probabilities and transition rate matrices of pure-birth (forward in time) or pure-death (backward in time) processes are derived using probabilistic arguments, without assuming stationarity. While most earlier derivations assume stationarity, already \citet{Grif98} noted that this is not necessary. Furthermore, we show that particle probabilities can also be obtained using orthogonal polynomials. 

\subsection{Particle models}

Conditional on a sample $(y,M)$ from the present time, bi-allelic mutation and drift models, allow for ancestral particle configurations $(i,m)$ with an ancestral sample size $m$, $0\leq m\leq M$, and an ancestral frequency of the focal allele $i$, $0\leq i\leq \min(y,m)$. Recall that, for a sample $(i,m)$, we use a hypergeometric likelihood with the Moran model and a binomial likelihood with the diffusion model, in both cases conditional on the allele proportion $x$ in the population. (A sample of size $m=0$ trivially has a likelihood of one.) Since both these likelihoods sum over the ordered states and thus contain a binomial coefficient, we require the probabilities of ordered states $\Pr^{*}(i\given m,y,M,\dots)$ for compatibility. Note that our considerations hold for both the decoupled Moran as well as the diffusion models, since the same particle transition probabilities and rates are obtained by sampling from Moran and diffusion models, as long as times are scaled appropriately, because the coalescent is consistent \citep{Ethe11}. In particular, the same beta-binomial sampling distribution results from both sampling schemes in equilibrium.

Given a sample of size $m$ and looking backward in time, there are $m(m-1)$ possible coalescent and $m$ possible mutation events  \citep[compare][]{Fais15}. The descendant state yields information on the ancestral state neither with a parent-independent mutation nor with a coalescent. Thus, both these events simply reduce the sample size by one. With the usual scaling of time, the rate of a coalescent is $m(m-1)$, that of a mutation $m\theta$, such that the rate of an event reducing the sample size by one is $m(m-1+\theta)$. With $i$ alleles of the focal type in the sample, the rates of possible events that reduce the number of focal alleles from $i$ to $i-1$ are: $i(i-1)$ for a coalescent and $\alpha\theta i$ for a mutation; the rates for an event that reduce the number of the other allelic type from $m-i$ to $m-i-1$ are: $(m-i)(m-i-1)$ for a coalescent and $\beta\theta(m-i)$ for a mutation. With the $N$-particle look-down process \citep{Donn96,Ethe11}, labels of particles in the population are ordered; at each coalescent, always the particle with the higher label disappears. Similarly, we assume that, irrespective whether a particle disappears via a coalescent or mutation event, always the particle with the higher label disappears. Thus, the ordering is independent of the allele configuration and the sequence of events. Let $\Pr^{*}(i\given m)$ indicate the probability of such an ordered sample $(i,m)$, with the relationship between the probabilities of unordered and ordered samples $\Pr(i\given m)=\binom{m}{i}\Pr^{*}(i\given m)$. 

Set the indicator variable $z_m=1$, if an allele of the focal type is removed (or added, looking forward in time), or $z_m=0$, otherwise. Conditional on an event removing (or adding, looking forward in time) a particle, the likelihood of obtaining the ordered sample $(i=y,m=M)$ from any of the ordered samples with $(i-z_m,m-1)$, irrespective of the ordering of the samples, is
\begin{equation}
\begin{split}
    &{\Pr}^*(y=i\given i-z_m,m-1,\alpha,\theta)=\\ &\qquad\frac{\big(i(i-1+\alpha\theta)\big)^{z_m}\big((m-i)(m-i-1+\beta\theta)\big)^{1-z_m}}{m(m-1+\theta)}\,.
\end{split}
\end{equation}
The likelihood of observing an unordered sample $(i=y,m=M)$ given a sample $(i-z_m,m-1)$ is then
\begin{equation}\label{eq:transition_particle}
\begin{split}
    \Pr(y=i\given i-z_m,m-1,\alpha,\theta)&=
    \frac{m}{i^{z_m}(m-i)^{1-z_m}}{\Pr}^*(i\given i-z_m,m-1,\alpha,\theta)\\
    &=\frac{(i-1+\alpha\theta)^{z_m}(m-i-1+\beta\theta)^{1-z_m}}{m-1+\theta}\,.
\end{split}
\end{equation}
Note that we did not assume stationarity, but conditioned on an event losing (looking backward) or creating (looking forward) a particle and the configuration immediately preceding it. During this infinitesimal interval, probabilities of coalescence and mutation must remain constant. Thus temporal changes in these parameters must be sufficiently smooth. More rigorously, the set of discontinuities in the scaled mutation rate must have measure zero. Otherwise, temporal changes in the population genetic parameters, can be accommodated.---The transition probability~(\ref{eq:transition_particle}) corresponds to that derived earlier \citep[][section 4.2]{Ethe09}, assuming stationarity. 

\subsection{Forward particle model}

Looking forward in time, once created, either by mutation or split, particles do not change their allele state. Conditional on a transition from $m$ to $m+1$, the transition probability from an unordered sample $(i,m)$ to an unordered sample $(i+z_m,m+1)$ is
\begin{equation}\label{eq:forward_transition}
        \Pr(z_{m}\given i,m,\alpha,\theta)=\frac{(i+\alpha\theta)^{z_{m}}(m-i+\beta\theta)^{1-z_{m}}}{m+\theta}\,.
\end{equation} 

Assuming stationarity, this sampling scheme corresponds to a generalization of the Polya urn model to real-valued starting proportions \citep{Grif98,Burd18}. The scheme can also be thought of as a bi-allelic version of Hoppe's urn \citep{Hopp84}, where with any new mutation, instead of a new color corresponding to a new allele, with probability $\alpha$ the focal and $(1-\alpha)$ the other allele is chosen. Under stationarity, the population allele proportion $x$ is beta-distributed with parameters $\alpha\theta$ and $(1-\alpha)\theta$. The initial $z_0$ is drawn from a Bernoulli with parameter equal to the expectation of the prior beta distribution of $x$, \ie\ $\E(x\given \alpha,\theta)=(\alpha\theta)/\theta=\alpha$. All further $z_{m}$ are drawn from a Bernoulli with parameter equal to the expectation of the posterior distribution of $x$ given $i$ and $m$, which in equilibrium is again a beta with parameters $\alpha\theta+i$ and $\beta\theta+m-i$, such that the expectation is $\E(x\given i,m,\alpha,\theta)=(i+\alpha\theta)/(m+\theta)$. From this scheme, the transition probability~(\ref{eq:forward_transition}) follows. For a sample of size $M$, the distribution resulting from this scheme is the beta-binomial
\begin{equation}\label{eq:beta_bin}
\begin{split}
    \Pr(y\given M,\alpha,\theta)&=\binom{M}{y}\frac{\Gamma(\theta)}{\Gamma(\alpha\theta)\Gamma(\beta\theta)}\frac{\Gamma(y+\alpha\theta)\Gamma((M-y)+\beta\theta)}{\Gamma(M+\theta)}\,.
\end{split}
\end{equation}

\subsection{Backward particle model}

Population genetic data are usually collected at the present time $t=0$, while past population genetic parameters are inferred. Demographic changes, \eg\ in the population size, happen at a certain time backward. Even if data are obtained at different time points, as is the case with evolve and re-sequence experiments \citep[\eg][]{Schl15} or with data from ancient samples \citep[\eg][]{Krause10}, computing the likelihood backwards is just as economic as forward \citep{Berg17}. As with the forward algorithm, a time-dependent part can be separated from a time-independent part. The following system of differential equations describes the time dependency: 
\begin{equation}\label{eq:temp_system_back}
\begin{split}
    -\frac{d}{dt}\Pr(m=M\given t,\theta, M)&=\lambda_M\,\Pr(m=M\given t,\theta, M)\,, \quad\text{and}\\ 
    -\frac{d}{dt}\Pr(m\given t,\theta, M)&=-\lambda_{m+1}\,\Pr(m+1\given t,\theta, M)\\
    &\quad+\lambda_{m}\,\Pr(m\given t,\theta, M)\,, \quad\text{for $M>m\geq 0$.}
\end{split}
\end{equation}
Obviously, the eigenvalues are $\lambda_m=m(m-1+\theta)$ with $0\leq m\leq M$. The starting condition is $\Pr(m=M\given t=0,\theta,M)=1$ and all other $\Pr(0\leq m <M\given t=0,\theta,M)=0$. Then the $\Pr(m\given t,\theta,M)$ are sums of exponential functions of $t$ (note that time is running backward, $0\leq t$):
\begin{equation}\label{eq:temp_system_back_solution}
\begin{split}
    \Pr(m=M\given t,\theta,M)&=e^{\lambda_M t}\,,\\
    \Pr(m\given t,\theta,M)&=\sum_{i=m}^{M}\frac{\prod_{j=m+1}^M \lambda_j}{\prod_{j=m,j\neq i}^M(\lambda_j-\lambda_i)}\,e^{\lambda_i t}\,,  \quad\text{for $M-1\geq m\geq 1$, and}\\
    \Pr(m=0\given t,\theta,M)&=1-\sum_{m=1}^M\Pr(m\given t,\theta, M)\,.
\end{split}
\end{equation}
This can be shown by substitution of the solution~(\ref{eq:temp_system_back_solution}) into the system of differential equations~(\ref{eq:temp_system_back}) (see Appendix~\ref{section:appendix_linear_diff}) and recursion.---Note that this solution was actually obtained via the equivalence of the solutions of the particle and the orthogonal polynomial approaches (see section \ref{section:allele_proportion_diffusion} and Appendix~\ref{section:appendix_diff_eqs}). 

The algorithm for the time-independent part is as follows for an ordered sample:
\begin{itemize}
    \item {Initiation.} Start with a sample of size $M$ at time $t=0$, where $y$ are of the focal type. Set $i=y$ and $m=M$. Introduce the vector of $(M+1)M/2$ backward variables $b^*(i,m)$, where $0\leq m\leq M$ and $0\leq i\leq m$. Set $b^*(i=y,m=M)=1$ and all other $b^*(i,m)=0$. The overall rate of moving from $m$ to $m-1$ is $m(m-1+\theta)$. 
    \item{Recursion.} Move from $m+1$ to $m$ by calculating for all $i$, with $0\leq i\leq m$
    \begin{equation}
    \begin{split}
     b^*(i,m)&= \frac{(m-i+1)(m-i+\beta\theta)}{(m+1)(m+\theta)}\,b^*(i,m+1)\\
     &\qquad+\frac{(i+1)(i+\alpha\theta)}{(m+1)(m+\theta)}\,b^*(i+1,m+1)\,.
    \end{split}
    \end{equation}    
    \item{Stop.} End the recursion, when the sample size $m=0$ is reached. 
\end{itemize}
The $b^*(i,m)$ can be interpreted as ${\Pr}^*(y\given M,i,m,\alpha,\theta)$; ${\Pr}^*(y\given M,i=0,m=0,\alpha,\theta)$ corresponds to the likelihood ${\Pr}^*(y\given M,\alpha,\theta)$. The algorithm for the probabilities of unordered samples follows analogously. In equilibrium, the probability of an unordered sample, \ie\  $\binom{M}{y}$ times the probability of an ordered sample, corresponds to the beta-binomial (\ref{eq:beta_bin}). The intermediate equilibrium probabilities are similar to the beta-binomial 
\begin{equation}\label{eq:cond_backw_particle}
\begin{split}
    {\Pr}(y\given M, i,m,\alpha,\theta)&=\binom{M-m}{y-i}
    \frac{\Gamma(m+\theta)}{\Gamma(i+\alpha\theta)\Gamma((m-i)+\beta\theta)}\\
    &\qquad\times\frac{\Gamma(y+\alpha\theta)\Gamma((M-y)+\beta\theta)}{\Gamma(M+\theta)}\,.
\end{split}
\end{equation}
The binomial term counts the number of ways to reach a configuration of $(y,M)$ from a configuration $(i,m)$. 

The transition probabilities from, forward-in-time $m=0$ to $m=3$ (or backward-in-time from $m=3$ down to $m=0$) are given in Table~(\ref{table:transitionp_reversible}). 

\begin{table}[ht]
\centering
\caption{A tabular form of the transition probabilities for the reversible mutation model from $m=0$ up to $m=3$.}
  \begin{tabular}{c|cccccc}
      $m$  &\multicolumn{6}{c}{$i$}\\
     &$0$ to $0$ &$0$ to $1$ &$1$ to $1$ &$1$ to $2$ &$2$ to $2$ &$2$ to $3$ \\
      \hline
    $2$ to $3$  &$\frac{3(2+\beta\theta)}{3(2+\theta)}$ &$\frac{\alpha\theta}{3(2+\theta)}$  &$\frac{2(1+\beta\theta)}{3(2+\theta)}$ &$\frac{2(1+\alpha\theta)}{3(2+\theta)}$   &$\frac{\beta\theta}{3(2+\theta)}$ &$\frac{3(2+\alpha\theta)}{3(2+\theta)}$\\  
     $1$ to $2$  &$\frac{2(1+\beta\theta)}{2(1+\theta)}$&$\frac{\alpha\theta}{2(1+\theta)}$   &$\frac{\beta\theta}{2(1+\theta)}$ &$\frac{2(1+\alpha\theta)}{2(1+\theta)}$\\
    $0$ to $1$ &$\frac{\beta\theta}{\theta}$ &$\frac{\alpha\theta}{\theta}$\\  
  \end{tabular}\label{table:transitionp_reversible}
\end{table}

\subsection{Population allele proportions using Jacobi polynomials}
\label{section:allele_proportion_diffusion}

Combining the time-dependent and time-independent parts from the previous subsection, we have backward in time,
\begin{equation}\label{eq:particle2polynomial}
\begin{split}
    \Pr(y\given M,x,\alpha,\theta,t)&=\sum_{m=0}^{M}{\Pr}(m\given t){\Pr}(y\given M, x, m, \alpha,\theta)\\
    &\text{with}\\
    \Pr(y\given M, x, m, \alpha,\theta)&=\sum_{i=\max(m+y-M,0)}^{\min(m,y)}{\Pr}^*(y\given M,i,m,\alpha,\theta)\, \binom{m}{i}x^i(1-x)^{m-i}\,.
\end{split}
\end{equation}
This is a polynomial of degree $M$ in $x$. 

The binomial likelihood $\Pr(y\given M,x,t=0)=\binom{M}{y}x^y(1-x)^{M-y}$ can also be expressed uniquely by an expansion in the modified Jacobi polynomials up to order $M$ (eq.~\ref{eq:binom2jacobi}), which we rewrite here: 
\begin{equation}
    \Pr(y\given M,x,t=0)=\binom{M}{y}x^y(1-x)^{M-y}=\sum_{m=0}^M d_m(M,y) R_m^{(\alpha,\theta)}(x)\,,
\end{equation}
where $d_m(M,y)$ is a constant depending on $m$, $M$, and $y$. Let $r_{m,l}$ be the coefficient of $R_m^{(\alpha,\theta)}(x)$ that is multiplied with the term $x^l$ and let $\mathbf{R}$ be the matrix of coefficients, a triangular matrix, which is only needed up to order $M$. Let $\mathbf{a}(M,y)$ be the coefficients of the polynomial expansion of the binomial likelihood. Then the vector $\mathbf{d}(M,y)$ solves the equation $\mathbf{R}\mathbf{d}(M,y)=\mathbf{a}(M,y)$. The $d_m(M,y)$ are also the solution of 
\begin{equation}
    d_m(M,y)=\frac{1}{\Delta_m^{(\alpha, \theta)}}\int_0^1 \binom{M}{y}x^y(1-x)^{M-y}\,R_m^{(\alpha,\theta)}(x)\, x^{\alpha-1}(1-x)^{\beta\theta-1}\,dx\,,
\end{equation}
with
\begin{equation}
    \Delta_m^{(\alpha,\theta)}=\frac{\Gamma(m+\alpha\theta)\Gamma(m+\beta\theta)}{(2m+\theta-1)\Gamma(m+\theta-1)\Gamma(m+1)}\,.
\end{equation}

At any time $t$ backward in time, we have
\begin{equation}\label{eq:orthopolynomial}
    \Pr(y\given M,x,\alpha,\theta,t)=\sum_{m=0}^M d_m(M,y) e^{\lambda_m t} R_m^{(\alpha,\theta)}(x)\,.
\end{equation}
Comparing equations~(\ref{eq:particle2polynomial}) and (\ref{eq:orthopolynomial}), we find that the two are identical. Thus we can equate terms multiplied by the same exponential function $e^{\lambda_m t}$ in eq.~(\ref{eq:particle2polynomial}) and the $d_m(M,y) R_m^{(\alpha,\theta)}(x)$ in eq.~(\ref{eq:orthopolynomial}). As the time-dependent dynamics depend only on the sample size $M$, but not on $y$, we have for $y=M$ (see Appendix~\ref{section:appendix_diff_eqs}),
\begin{equation}
    d_m(M,M) r_{m,m}=c_{M,m} b_{m,m}^{*}\,,
\end{equation}
where $c_{M,m}$ is the $m$th coefficient solving the system of temporal differential equations (\ref{eq:temp_system_back}). Using the fact that the backward variables $b^*(i,m)={\Pr}^*(y\given M, i,m,\alpha,\theta)$ are associated with the $m$th eigenvalue, we can use eq.~(\ref{eq:cond_backw_particle}) to recover the $b^*(i,m)$ from the modified Jacobi expansion:
\begin{equation}
    \sum_{i=\max(m+y-M,0)}^{\min(m,y)}{\Pr}^*(y\given M, i,m,\alpha,\theta)= \frac{d_m(M,y) r_{m,m}}{c_{M,m}} \,.
\end{equation}
Since the polynomials are eigenfunctions of the backward diffusion equation and eq.~(\ref{eq:cond_backw_particle}) is relatively simple, particle probabilities can be calculated efficiently. 

Assuming equilibrium, the marginal likelihood corresponds to the zeroth coefficient $\Pr(y\given M,x,\alpha,\theta)=d_0(M,y)$. With the coalescence approach, this result would have been obtained only after a lengthy recursion from the observed sample $(y,M)$ down to the empty sample $(0,0)$.

\section{Conditional probability of ancestral particle configurations}

In population genetics, population demographic events usually are modeled to occur at a specific time in the past. We illustrate with a simple model, where i) at time $t=0$, sample data $(y,M)$ are given, ii) between the time $0<t<s$, the population genetic parameters are $\alpha$ and $\theta$, and iii) at time $t=s$ in the past the allele proportion is a beta $\rho(x,t=s)=\frac{\Gamma(\vartheta)}{\Gamma(\alpha\vartheta)\Gamma(\beta\vartheta)}\,x^{\alpha\vartheta-1}(1-x)^{\beta\vartheta-1}$. With a change in the effective population size at time $t=s$, $\theta$ will differ from $\vartheta$, while $\beta$ is assumed constant throughout. Assume that we want to calculate at a specific time $t$, $0<t<s$, in the past, the joint probability
$\Pr(y,i,m \given M,t,\alpha,\theta)$ and the conditional probability $\Pr(i,m \given y,M,t,\alpha,\theta)$. If $\theta=\vartheta$, \ie\ in equilibrium, $\Pr(i\given m,t,\alpha,\theta)$ would correspond to a beta-binomial. 

\subsection{Example: conditional probability under equilibrium}

\begin{figure}[ht]
    \centering
    \includegraphics[width = 11.5cm]{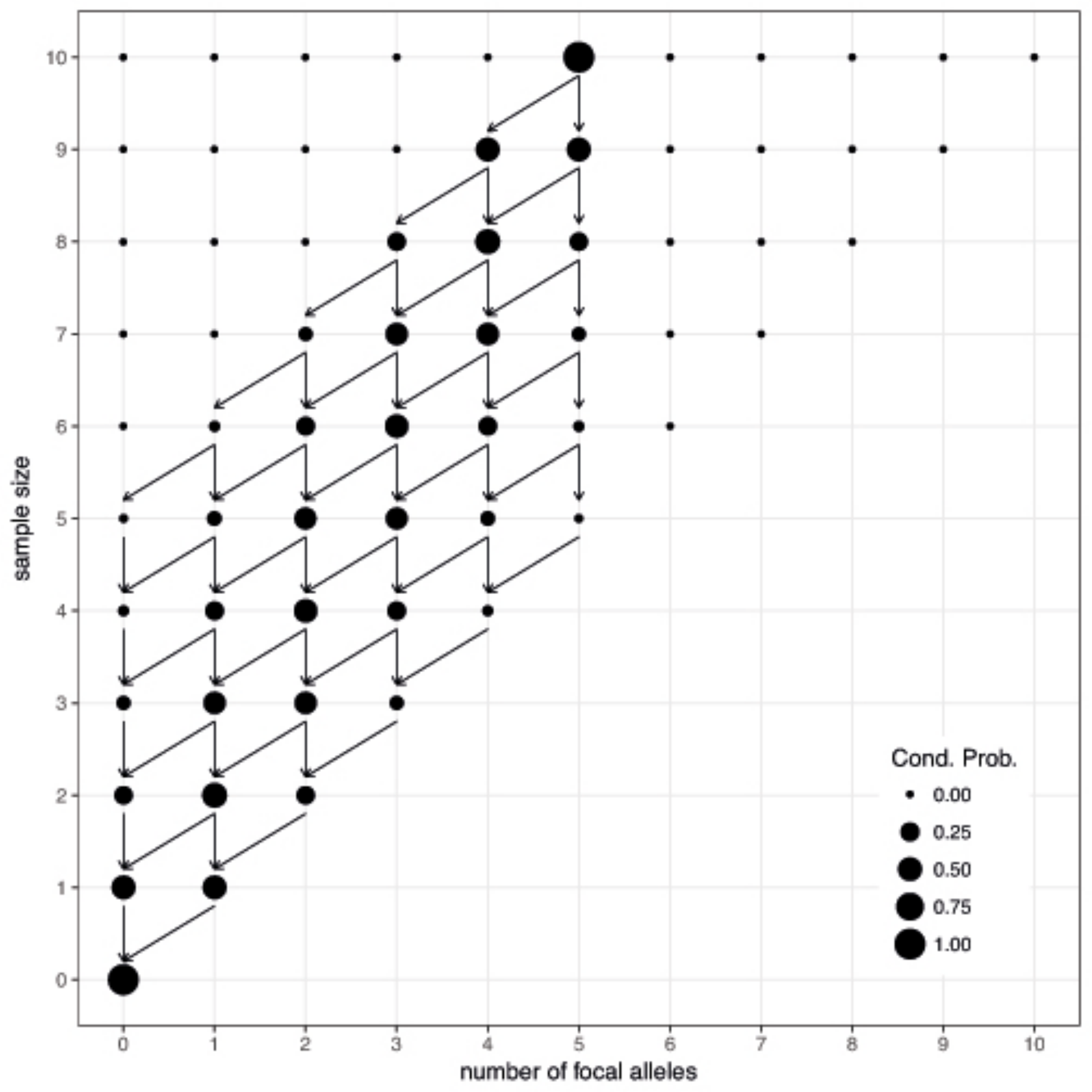}
    \caption{Conditional probability of ancestral particle configurations under equilibrium for a sample of size $M = 10$. At the beginning we assume the presence of $y = 5$ focal alleles. The size of the dots represents the conditional probability of the particular configuration. Arrows point to all possible configurations backwards in time for each pair $(y,M)$.}
    \label{fig:num_example_equilibrium}
\end{figure}

We now calculate the conditional probabilities in a numerical example under equilibrium. Consider a sample of size $M = 10$ with five focal alleles, \ie\ a starting configuration of $(5,10)$. Assume a mutation bias towards the focal allele of $\alpha = 0.3$ and a scaled mutation rate of $\theta = 0.1$. Fig.~\ref{fig:num_example_equilibrium} shows the conditional probability of ancestral particle configurations including all possible paths going backwards in time from $(5,10)$ to $(0,0)$.

\subsection{Conditional probability under non-equilibrium}

With the backward particle model, it is possible to calculate for each sample $(i,m)$ at time $t$ the probability of $\Pr(i\given m,t,\alpha,\theta)$ even with changing population sizes, \ie\ $\theta\neq \vartheta$. As many intermediate configurations $(i,m)$ are possible, this strategy will generally be computationally demanding. With the forward particle model, the sample size $m$ is a random variable conditional on population parameters and the starting time of the process, if stationarity is not assumed. As a workaround, one could start at $t=s$ with a sample of size $M$ from $\rho(x)$ and then use the hypergeometric distribution to obtain samples of sizes $1\leq m\leq M$, for all $t<0$. But as the sample sizes to be considered are bigger than necessary, this strategy is computationally wasteful. This can be avoided by using orthogonal polynomials. 

Expand the beta $(\alpha,\theta)$ distribution into modified Jacobi polynomials with parameters $\theta$ and $\alpha$ to obtain
\begin{equation}
\begin{split}
    \phi(x\given \theta,\alpha,t,\rho)&=\frac{\Gamma(\vartheta)}{\Gamma(\alpha\vartheta)\Gamma(\beta\vartheta)}\,x^{\alpha\vartheta-1}(1-x)^{\beta\vartheta-1}\\
    &=\sum_{n=0}^\infty \rho_n R_n^{(\alpha,\theta)}(x)  e^{\lambda_n(s-t)} x^{\alpha\theta-1}(1-x)^{\beta\theta-1}\,.
\end{split}
\end{equation}
From the particle approach, we obtain
\begin{equation}
    \Pr(y\given M,i,m,\alpha,\theta)\,.
\end{equation}
Expand the binomial likelihood of $(i,m)$ at time $t$ into modified Jacobi polynomials
\begin{equation}
    \Pr(i\given m,x)=\binom{m}{i} x^{i}(1-x)^{m-i}=\sum_{j=0}^m \check{d}_j(m,i) R_j^{(\alpha,\theta)}(x)\,.
\end{equation}
The joint probability of the ancestral and the extant particle configuration at time $t$, then is 
\begin{equation}\label{eq:joint_y_i}
\begin{split}
    &\Pr(y,i,m\given M,m,\alpha,\theta,t,\rho)\\
    &\qquad=\int_0^1 \Pr(y\given M,i,m,\alpha,\theta)\,\Pr(m\given M,\theta,t)\,\Pr(i\given m,x)\,\phi(x\given \alpha,\theta,t,\rho)\,dx\\
    &\qquad=\Pr(y\given M,i,m,\alpha,\theta)\,\Pr(m\given M,\theta,t)\,e^{\lambda_j(s-t)}\\
    &\qquad\qquad \times\sum_{j=0}^m\int_0^1 \check{d}_j(m,i)  R_j^{(\alpha,\theta)}(x) \rho_j R_j^{(\alpha,\theta)}(x)
     x^{\alpha\theta-1}(1-x)^{\beta\theta-1}\,dx\\
   &\qquad= \Pr(y\given M,i,m,\alpha,\theta)\,\Pr(m\given M,\theta,t)\,\sum_{j=0}^m \check{d}_j(m,i)\rho_j \Delta_j^{(\alpha,\theta)} e^{\lambda_j(s-t)}\,.
\end{split}
\end{equation}
Dividing this joint likelihood by the marginal likelihood $\Pr(y\given M,\alpha,\theta,t,\rho)$, we get the conditional likelihood $\Pr(i,m\given y,M,\alpha,\theta,t,\rho)$. This strategy is computationally relatively easy: only an expansion of $\rho$ up to $M$ is needed, and with the single forward expansion, all configurations $i$, $0\leq i\leq m$, and $m$, $0\leq m\leq M$ can be covered. This is numerically simpler than any of the particle approaches presented earlier.

Furthermore, as is evident from eq.~(\ref{eq:joint_y_i}), the joint probability of the configuration $(y,M)$ at time $0$ and $(i,m)$ at time $t$, $\Pr(y,i\given M,m,\alpha,\theta,t,\rho)$, is independent of $x$. The same holds true for the conditional probability of $(i,m)$ given $(y,M)$, $\Pr(i,m\given y,M,\alpha,\theta,t,\rho)$.

\subsection{Example: Non-equilibrium caused by a change in the scaled mutation rate}

\begin{figure}[ht]
    \centering
    \includegraphics[width = 11.5cm]{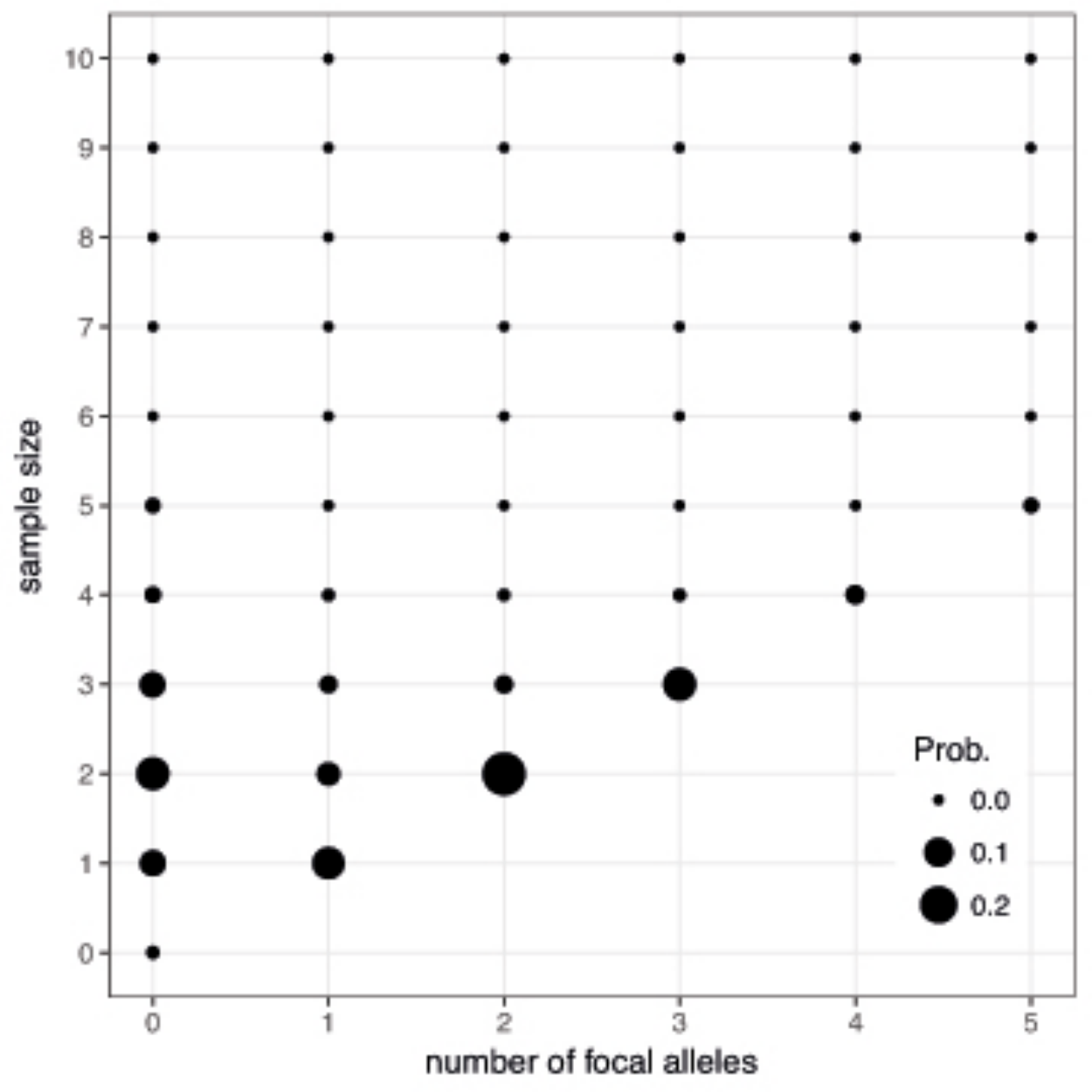}
    \caption{Conditional probability of ancestral particle configurations under non-equilibrium for a sample of size $M = 10$. At time $t=0$, we assume the presence of $y = 5$ focal alleles. The size of the dots represents the conditional probability of the particular configuration at time $t=s=0.5$.}
    \label{fig:num_example_non-equilibrium}
\end{figure}

Assume a single change in the scaled mutation rate at time $s=-0.5$ from $\theta_a=0.3$ in the past to $\theta_c=0.1$ from then on; $\alpha=0.3$ for all previous times, such that the distribution is in equilibrium at $t=s=-0.5$. The data are $y=5$ with a sample size of $M=10$ at time $t=0$. At time $t=s$ in the past, we thus have the beta-binomial~(\ref{eq:beta_bin}) with $i$ replacing $y$, $m$ replacing $M$, and $\theta_a$ replacing $\theta$. The probabilities of $m$ given $t=s$ are given by equation~(\ref{eq:temp_system_back_solution}). The probability of $y$ given $i$, $m$, and $M$ is given by equation~(\ref{eq:cond_backw_particle}) with $\theta_c$ replacing $\theta$. Altogether, we thus have for $1\leq m\leq (M-1)$:
\begin{equation}\label{eq:joint_y_i_at_s}
\begin{split}
    &\Pr(y,i,m\given M,m,\alpha,\theta,t,\rho)=\Pr(y\given M,i,m,\alpha,\theta_c)\Pr(m\given M,\theta_c,t)\Pr(i\given m,\alpha,\theta_a)\\
    &\qquad=\binom{M-m}{y-i} \frac{\Gamma(m+\theta_c)}{\Gamma(i+\alpha\theta_a)\Gamma((m-i)+\beta\theta_c)}\frac{\Gamma(y+\alpha\theta_c)\Gamma((M-y)+\beta\theta_c)}{\Gamma(M+\theta_c)}\\
    &\qquad\qquad\times\sum_{j=m}^{M}\frac{\prod_{k=m+1}^M \lambda_k}{\prod_{k=m,k\neq j}^M(\lambda_k-\lambda_j)}\,e^{\lambda_j t}\\
    &\qquad\qquad\times\binom{m}{i}\frac{\Gamma(\theta_a)}{\Gamma(\alpha\theta_a)\Gamma(\beta\theta_a)}\frac{\Gamma(i+\alpha\theta_a)\Gamma((m-i)+\beta\theta_a)}{\Gamma(m+\theta_a)}\,
\end{split}
\end{equation}
with the eigenvalues $\lambda_j=j(j-1+\theta_c)$. The probabilities of the various configurations $i$ and $m$ are given in Fig.~(\ref{fig:num_example_non-equilibrium}).

\section{Summary}

Initially, we reviewed a biallelic mutation-drift decoupled Moran model, the diffusion model equations, as well as the particle model (which corresponds to the coalescent backward in time) and how to calculate joint and conditional probabilities, given a small or moderate sized sample from the present. Usually in population genetics, we use such samples to infer parameters of past population processes. If only the marginal likelihood is desired, a backward model with orthogonal polynomials is simple and numerically efficient under these assumptions \citep{Berg17}. With the bi-allelic mutation-drift model, inference is not confined to equilibrium conditions, but also allows for changing population sizes, as long as the discontinuities are a set of measure zero. Population genetic parameters, \eg\ a changing effective population size or mutation bias, can be inferred relatively easily compared to particle approaches. If the probability of an ancestral configuration some time in the past is desired, these backward-in-time approaches need to be combined with forward-in-time approaches. If additionally changing population sizes are assumed, the sample size becomes a random variable for forward-in-time particle processes. This makes particle-based approaches, \ie\ the coalescence and the forward-in-time particle model, numerically inefficient. Orthogonal polynomials, however, allow for efficient numerical calculations, as we show in this article. When deriving the equations, relationships between the solutions of the diffusion equation with orthogonal polynomials and particle models (in backward-time corresponding to the coalescent) are established that show details of their dual relationship.

\section{Appendices}

\subsection{Solution of the linear system of differential equations}\label{section:appendix_linear_diff}

The linear system of differential equations~(\ref{eq:temp_system_back}) is solved by the vector~(\ref{eq:temp_system_back_solution}) as can be shown by substitution. For $m=M$, it is obvious that $\Pr(m=M\given t,\theta,M)=1-e^{\lambda_M t}$ solves $-\frac{d}{dt}\Pr(m=M\given t,\theta,M)=M(M-1+\theta)\,\Pr(m=M\given t,\theta,M)$. For $M-y\geq m \geq 1$, the left hand side is:
\begin{equation}
\begin{split}
        -\frac{d}{dt}\Pr(m\given t,\theta,M) &=\sum_{i=m}^{M}\frac{\lambda_i\,\prod_{j=m+1}^M \lambda_j}{\prod_{j=m,j\neq i}^M(\lambda_j-\lambda_i)}\,e^{\lambda_i t}\,.
\end{split}
\end{equation}
This is equal to the right hand side:
\begin{equation}
\begin{split}
        &-(m+1)(m+\theta)\,\Pr(m+1\given t,\theta,M)+m(m-1+\theta)\,\Pr(m+1\given t,\theta,M)\\
        &\qquad =-\sum_{i=m+1}^{M}\frac{\lambda_{m+1}\,\prod_{j=m+2}^M \lambda_j}{\prod_{j={m+1},j\neq i}^M(\lambda_j-\lambda_i)}\,e^{\lambda_i t}+\sum_{i=m}^{M}\frac{\lambda_{m}\,\prod_{j=m+1}^M \lambda_j}{\prod_{j=m,j\neq i}^M(\lambda_j-\lambda_i)}\,e^{\lambda_i t}\\
        &\qquad =-\sum_{i=m+1}^{M}\frac{\prod_{j=m+1}^M \lambda_j}{\prod_{j={m+1},j\neq i}^M(\lambda_j-\lambda_i)}\,e^{\lambda_i t}\\
        &\qquad\qquad+\sum_{i={m+1}}^{M}\frac{\lambda_{m}\,\prod_{j=m+1}^M \lambda_j}{(\lambda_m-\lambda_i)\prod_{j={m+1},j\neq i}^M(\lambda_j-\lambda_i)}\,e^{\lambda_i t}-\frac{\lambda_{m}\,\prod_{j=m+1}^M \lambda_j}{\prod_{j=m+1}^M(\lambda_j-\lambda_m)}\,e^{\lambda_m t}\\
        &\qquad =-\sum_{i=m+1}^{M}\frac{((\lambda_m-\lambda_i)-\lambda_m)\,\prod_{j=m+1}^M \lambda_j}{(\lambda_m-\lambda_i)\prod_{j={m+1},j\neq i}^M(\lambda_j-\lambda_i)}\,e^{\lambda_i t}+\frac{\lambda_{m}\,\prod_{j=m+1}^M \lambda_j}{\prod_{j=m+1}^M(\lambda_j-\lambda_m)}\,e^{\lambda_m t}\\
        &\qquad =\sum_{i=m}^{M}\frac{\lambda_i\,\prod_{j=m+1}^M \lambda_j}{\prod_{j={m},j\neq i}^M(\lambda_j-\lambda_i)}\,e^{\lambda_i t}\,.
\end{split}
\end{equation}
The last term for $m=0$, is one minus the sum of all other terms, \ie\  
$$\Pr(m=0\given t,\theta,M)=1-\sum_{m=1}^M\Pr(m\given t,\theta, M)\,.$$

\subsection{Appendix: Solution of backward systems of ordinary time-dependent differential equations}\label{section:appendix_diff_eqs}


Start with the system (\ref{eq:temp_system_back}), rewritten here with ${\phi}_m(t)$ and $\lambda_m$:
\begin{equation}\label{eq:temp_system_back_alt}
\begin{split}
    -\frac{d}{dt}{\phi}_m(t)=\lambda_m\,{\phi}_m(t) &\qquad\text{for $m=M$,}\\
    -\frac{d}{dt}{\phi}_m(t)=\lambda_{m}\,{\phi}_{m}(t)+\lambda_{m+1}\,{\phi}_{m+1}(t) &\qquad\text{for $0\geq m\geq (M-1)$}\,.
\end{split}
\end{equation}
The starting condition is ${\phi}_M(t=0)=1$ and all other ${\phi}_m(t=0)=0$ equal zero. Note that ${\phi}_1(t)=\sum_{i=2}^{M}{\phi}_m(t)$. We therefore leave ${\phi}_0(t)$ and $\lambda_0=0$ from the system. Consider the corresponding transition rate matrix
\begin{equation}
    \mathbf{L}=\begin{pmatrix}
    -\lambda_M &\lambda_M &0 &\hdots  &0 &0 &0\\
    0 &-\lambda_{M-1} &\lambda_{M-1} &\hdots &0 &0 &0\\
    \vdots &\vdots &\ddots &\ddots &\vdots &\vdots  &\vdots \\
    \vdots &\vdots &\vdots &\ddots &\ddots &\vdots &\vdots\\
    0 &0 &0 &\hdots &-\lambda_2 &\lambda_2 &0\\
    0 &0 &0 & \hdots  &0 &-\lambda_1 &\lambda_1\\
    \end{pmatrix}
\end{equation}
Because the matrix is triangular, the eigenvalues are the elements on the main diagonal, $\mathbf{\lambda}=(-\lambda_M,-\lambda_{M-1},\dots,-\lambda_2,-\lambda_1)$. 

The left eigenvector equations are 
\begin{equation}
    \mathbf{u}_m\mathbf{L}=-\lambda_m \mathbf{u}_m\,.
\end{equation}
For the eigenvector with eigenvalue $-\lambda_M$, we substitute 
\begin{equation}
\mathbf{u}_M=(1,u_{M,M},u_{M,M-1},\dots,u_{M,1})\,,
\end{equation}
to get
\begin{equation}
\begin{split}
    -\lambda_M&=-\lambda_M\\
    \lambda_M-\lambda_{M-1} u_{M,M-1}&=-\lambda_M u_{M,M-1}\\
    \lambda_{M-1}u_{M,M-1}-\lambda_{M-2} u_{M,M-2}&=-\lambda_Mu_{M,M-2}\\
    \vdots\qquad &=\quad \vdots\\
    \lambda_2u_{M,2}-\lambda_1 u_{M,1}&=-\lambda_{M}u_{M,1}\\
    \end{split}
\end{equation}
This leads to the recursion $u_{M,M}=1$ and for $1 \geq m \geq (M-1)$ 
\begin{equation}
    u_{M,m}=\frac{\lambda_{m+1}}{\lambda_m-\lambda_M}u_{M,m+1}\,.
\end{equation}
The solution to the recursion is: $u_{M,M}=1$ and
\begin{equation}
    u_{M,m}=\frac{\prod_{j={m+1}}^{M}\lambda_j}{\prod_{j=m}^{M-1}(\lambda_j-\lambda_M)}\,.
\end{equation}

Generally, for the eigenvalue $-\lambda_n$, we substitute 
\begin{equation}
\mathbf{u}_n=(u_{M,M},u_{M,M-1},\dots,u_{n,n+1},1,u_{n,n-1},\dots,u_{M,1})\,.    
\end{equation}
We then get
\begin{equation}
    -\lambda_M u_{n,M}=-\lambda_n u_{n,M}
\end{equation}
from which we deduce that $u_{n,M}=0$. For $m>n$, the next equations are    
\begin{equation}
    -\lambda_{m} u_{n,m}=-\lambda_n u_{n,m}
\end{equation}
from which we again deduce that $u_{n,m}=0$. From equation $n$ onward, we have  
\begin{equation}
\begin{split}
    -\lambda_n&=-\lambda_n\\
    \lambda_n-\lambda_{n-1} u_{n,n-1}&=-\lambda_n u_{n,n-1}\\
    \lambda_{n-1}u_{n,n-1}-\lambda_{n-2} u_{n,n-2}&=-\lambda_Mu_{n,n-2}\\
    \vdots\qquad &=\quad \vdots\\
    \lambda_2u_{n,2}-\lambda_1 u_{n,1}&=-\lambda_{n}u_{n,1}\\
    \end{split}
\end{equation}
This leads to the recursion: $u_{n,m}=0$ for $m>n$, $u_{n,n}=1$, and for $m<n$
\begin{equation}
    u_{n,m}=\frac{\lambda_{m+1}}{\lambda_n-\lambda_m}u_{n,m+1}\,.
\end{equation}
The solution to the recursion is: $u_{n,m>n}=0$, $u_{n,n}=1$, and for $n>m$
\begin{equation}
    u_{n,m}=\frac{\prod_{j=m+1}^{n}\lambda_j}{\prod_{j=m}^{n-1}(\lambda_j-\lambda_n)}\,.
\end{equation}

Our starting conditions at $t=0$ are $({\phi}_M(t),{\phi}_{M-1}(t),\dots,{\phi}_2(t),{\phi}_1(t))=(1,0,\dots,0,0)$. The coefficients fulfilling the starting condition are $c_{M,M}=1$, 
\begin{equation}
    c_{M,M-1}=-u_{M,M-1}=\frac{\lambda_M}{\lambda_{M}-\lambda_{M-1}}\,,
\end{equation}
\begin{equation}
\begin{split}
    c_{M,M-2}&=-u_{M,M-2}-c_{M,M-1}u_{M-1,M-2}\\
    &=-\frac{\lambda_M\lambda_{M-1}}{(\lambda_{M-1}-\lambda_M)(\lambda_{M-2}-\lambda_M)}-\frac{\lambda_M}{\lambda_M-\lambda_{M-1}}\frac{\lambda_{M-1}}{\lambda_{M-2}-\lambda_{M-1}}\\
    &=\frac{\lambda_M\lambda_{M-1}(\lambda_{M-2}-\lambda_{M-1}-\lambda_{M-2}+\lambda_M)}{(\lambda_M-\lambda_{M-1})(\lambda_M-\lambda_{M-2})(\lambda_{M-1}-\lambda_{M-2})}\\
    &=\frac{\lambda_M\lambda_{M-1}}{(\lambda_M-\lambda_{M-2})(\lambda_{M-1}-\lambda_{M-2})}\,.
\end{split}
\end{equation}
Generally, 
\begin{equation}\label{eq:c_m}
    c_{M,m}=\frac{\prod_{i=m+1}^{M}\lambda_i}{\prod_{i=m+1}^{M}(\lambda_i-\lambda_m)}\,,
\end{equation}
which we will proof below. Eventually, we have
\begin{equation}
\begin{split}
    &({\phi}_M(t),{\phi}_{M-1}(t),\dots,{\phi}_2(t),{\phi}_1(t))\\
    &\qquad=\mathbf{u}_M e^{-\lambda_M t} +c_{M,M-1}\mathbf{u}_{M-1}e^{-\lambda_{M-1} t}
    +\dots
    +c_{M,2}\mathbf{u}_2e^{-\lambda_2 t}
    +c_{M,1}\mathbf{u}_1e^{-\lambda_1 t}\,.
\end{split}
\end{equation}

\subsubsection{Proof of formula~(\ref{eq:c_m})} 

Note that the length of the eigenvectors $\mathbf{u}_m$ corresponds to their order $m$ and that the $c_{M,m}$ in formula~(\ref{eq:c_m}) multiply the highest term $u_{m,m}$. For proof of formula~(\ref{eq:c_m}), we will use the equivalence of $\Pr(m=M\given M,\alpha,\theta,t)$ with the coalescence and the modified Jacobi polynomials (see section~\ref{section:allele_proportion_diffusion}). As the temporal part is independent of the particular starting configuration, a monomorphic sample of sample size $M$, say $\Pr(m=M\given M,\alpha,\theta,t)$, may be chosen:
\begin{equation}
\begin{split}
    b^{*}(m,m)&=\frac{\prod_{i=m}^M(i-1+\alpha\theta)}{\prod_{i=m}^M(i-1+\theta)}\\
    &=\frac{\Gamma(m+\theta)}{\Gamma(M+\theta)}\frac{\Gamma(M+\alpha\theta)}{\Gamma(m+\alpha\theta)}\,.
\end{split}
\end{equation}
The temporal evolution of $\Pr(x\given M,\alpha,\theta,t)$ with the particle approach is
\begin{equation}
    \Pr(x\given M,\alpha,\theta,t)=\sum_{m=1}^M {\phi}_m(t)\,b^{*}(m,m)\,x^m.
\end{equation}
A binomial likelihood of a site occupying a specific polymorphic state $y$ can be written as a polynomial
\begin{equation}\label{eq:blik}
\Pr(y\given x,M,\alpha,\theta,t=0)= \sum_{m=0}^M a_m(M,y) x^m= \sum_{m=0}^M d_m(M,y) R_m^{(\alpha,\theta)}(x)\,,
\end{equation}
where $a_m(M,y)$ are the coefficients of the expansion in ordinary polynomials, $R_m^{(\alpha, \theta)}$ are the modified Jacobi polynomials of degree $m$ with base $(\alpha, \theta)$, 
\begin{equation}
  R_m^{(\alpha,\theta)}(x)=\sum_{l=0}^n(-1)^l\frac{\Gamma(m-1+l+\theta)\Gamma(m+\alpha\theta)}{\Gamma(m-1+\theta)\Gamma(l+\alpha\theta)l!(m-l)!}x^l\,,
\end{equation}
while $d_m(M,y)$ are coefficients for expanding the binomial likelihood in terms of $R_m^{(\alpha, \theta)}$. The $d_m(M,y)$ are also the solution of 
\begin{equation}
    d_m(M,y)=\frac{1}{\Delta_m^{(\alpha, \theta)}}\int_0^1 \Pr(y\given M,\alpha,\theta)\,R_m^{(\alpha,\theta)}(x)\, x^{\alpha-1}(1-x)^{\beta\theta-1}\,dx\,,
\end{equation}
with
\begin{equation}
    \Delta_m^{(\alpha,\theta)}=\frac{\Gamma(m+\alpha\theta)\Gamma(m+\beta\theta)}{(2m+\theta-1)\Gamma(m+\theta-1)\Gamma(m+1)}\,.
\end{equation}

Let $r_{m,l}$ be the coefficient of $R_m^{(\alpha,\theta)}(x)$ that is multiplied with the term $x^l$. For $y=M$, we have
\begin{equation}
    d_m(M,M) r_{m,m}=c_{M,m} b_{m,m}^{*}\,,
\end{equation}

\paragraph{Mellin Transform}
According to http://dlmf.nist.gov/18.17.36, we have
\begin{equation}
    \int_{-1}^{1} (1-y)^{z-1}(1+y)^a P_m^{(a,b)}(y)\,dy=
    \frac{2^{a+z}\Gamma(z)\Gamma(1+a+m)(1+b-z)_m}{m!\Gamma(1+a+z+m)}\,.
\end{equation}
Substituting $M-b$ for $z-1$ and then setting $R_m^{\alpha,\theta}(x)=P_m^{\beta\theta-1,\alpha\theta-1}(2x-1)$, we get
\begin{equation}
\begin{split}
    \int_{0}^{1} x^{M+\alpha\theta-1}(1-x)^{\beta\theta-1} R_m^{(\alpha,\theta)}(x)\,dx&=
    \frac{\Gamma(M+\alpha\theta)\Gamma(\beta\theta +m)\Gamma(M+1)}{m!\Gamma(\theta+M+m)\Gamma(M-m+1)}\\
    &=\frac{M!}{m!(M-m)!}\frac{\Gamma(\alpha\theta+M)\Gamma(\beta\theta +m)}{\Gamma(\theta+M+m)}\,.
\end{split}
\end{equation}
Therefore, 
\begin{equation}
\begin{split}
    c_{M,m}&= \frac{d_m(M,M) r_{m,m}}{b_{m,m}^{*}}\\
    &=\frac{(2m+\theta-1)\Gamma(m+\theta-1)\Gamma(m+1)}{\Gamma(m+\alpha\theta)\Gamma(m+\beta\theta)} \frac{M!}{m!(M-m)!}\frac{\Gamma(\alpha\theta+M)\Gamma(\beta\theta +m)}{\Gamma(\theta+M+m)}\\
    & \qquad\frac{\Gamma(2m-1+\theta)\Gamma(m+\alpha\theta)}{\Gamma(m-1+\theta)\Gamma(m+\alpha\theta)m!} \frac{\Gamma(M+\theta)}{\Gamma(m+\theta)}\frac{\Gamma(m+\alpha\theta)}{\Gamma(M+\alpha\theta)}\\
    &=\frac{M!}{m!(M-m)!}\frac{\Gamma(2m+\theta)}{\Gamma(\theta+M+m)}
    \frac{\Gamma(M+\theta)}{\Gamma(m+\theta)}\\
    &=\frac{\prod_{i=m+1}^M \lambda_i}{\prod_{i=m+1}^M (\lambda_i-\lambda_m)}\,,
\end{split}
\end{equation}
where the last line follows from the previous since we have, for $1< l<m$,
\begin{equation}\label{eq:eigval_diff}
\begin{split}
    \lambda_m-\lambda_l&=m(m-1+\theta)-l(l-1+\theta)\\
    &=(m-l)(m+l-1+\theta)\,,
\end{split}
\end{equation}
such that
\begin{equation}
\begin{split}
    c_{M,m}
    &=\frac{\prod_{i=m+1}^{M}\lambda_i}{\prod_{i=m+1}^{M}(\lambda_i-\lambda_m)}\\
    &=\frac{\prod_{i=m+1}^{M}i(i-1+\theta)}{\prod_{i=m+1}^{M} (i-m)(m+i-1+\theta)}\\
    &=\frac{M!}{(M-m)!m!}\frac{\Gamma(M+\theta)\Gamma(2m+\theta)}{\Gamma(m+\theta)\Gamma(m+M+\theta)}\,.
\end{split}
\end{equation}

\section*{Acknowledgments}

We thank the present and past members of the doctorate college population genetics, especially Juraj Bergman, for stimulating discussions. Our research was supported by the Austrian Science Fund (FWF): DK W1225-B20.

\section*{References}

\bibliography{coal}

\end{document}